\documentclass[nojss]{jss}\usepackage[]{graphicx}\usepackage[]{xcolor}
\makeatletter
\def\maxwidth{ %
  \ifdim\Gin@nat@width>\linewidth
    \linewidth
  \else
    \Gin@nat@width
  \fi
}
\makeatother

\definecolor{fgcolor}{rgb}{0.345, 0.345, 0.345}
\newcommand{\hlnum}[1]{\textcolor[rgb]{0.686,0.059,0.569}{#1}}%
\newcommand{\hlstr}[1]{\textcolor[rgb]{0.192,0.494,0.8}{#1}}%
\newcommand{\hlcom}[1]{\textcolor[rgb]{0.678,0.584,0.686}{\textit{#1}}}%
\newcommand{\hlopt}[1]{\textcolor[rgb]{0,0,0}{#1}}%
\newcommand{\hlstd}[1]{\textcolor[rgb]{0.345,0.345,0.345}{#1}}%
\newcommand{\hlkwa}[1]{\textcolor[rgb]{0.161,0.373,0.58}{\textbf{#1}}}%
\newcommand{\hlkwb}[1]{\textcolor[rgb]{0.69,0.353,0.396}{#1}}%
\newcommand{\hlkwc}[1]{\textcolor[rgb]{0.333,0.667,0.333}{#1}}%
\newcommand{\hlkwd}[1]{\textcolor[rgb]{0.737,0.353,0.396}{\textbf{#1}}}%

\usepackage{framed}
\makeatletter
\newenvironment{kframe}{%
 \def\at@end@of@kframe{}%
 \ifinner\ifhmode%
  \def\at@end@of@kframe{\end{minipage}}%
  \begin{minipage}{\columnwidth}%
 \fi\fi%
 \def\FrameCommand##1{\hskip\@totalleftmargin \hskip-\fboxsep
 \colorbox{shadecolor}{##1}\hskip-\fboxsep
     \hskip-\linewidth \hskip-\@totalleftmargin \hskip\columnwidth}%
 \MakeFramed {\advance\hsize-\width
   \@totalleftmargin\z@ \linewidth\hsize
   \@setminipage}}%
 {\par\unskip\endMakeFramed%
 \at@end@of@kframe}
\makeatother

\definecolor{shadecolor}{rgb}{.97, .97, .97}
\definecolor{messagecolor}{rgb}{0, 0, 0}
\definecolor{warningcolor}{rgb}{1, 0, 1}
\definecolor{errorcolor}{rgb}{1, 0, 0}
\newenvironment{knitrout}{}{} 

\usepackage{alltt}

\usepackage{orcidlink,thumbpdf,lmodern}

\usepackage{bm} 
\usepackage{amsmath} 
\usepackage{booktabs} 
\usepackage[color=blue!20!white,textsize=tiny,textwidth=22mm]{todonotes} 
\usepackage{mathtools}
\mathtoolsset{showonlyrefs}
\usepackage{longtable}

\newcommand{\eg}{\emph{e.g.,}}

\author{Emma Skarstein~\orcidlink{0000-0002-6347-855X}\\Norwegian University of\\Science and Technology
   \And Stefanie Muff\\Norwegian University of\\Science and Technology}
\Plainauthor{Emma Skarstein, Stefanie Muff}

\title{\pkg{inlamemi}: An \proglang{R} package for missing data imputation and measurement error modelling using INLA}
\Plaintitle{inlamemi: An R package for missing data imputation and measurement error modelling using INLA}
\Shorttitle{\pkg{inlamemi}}

\Abstract{
Measurement error and missing data in variables used in statistical models are common, and can at worst lead to serious biases in analyses if they are ignored. Yet, these problems are often not dealt with adequately, presumably in part because analysts lack simple enough tools to account for error and missingness. In this \proglang{R} package, we provide functions to aid fitting hierarchical Bayesian models that account for cases where either measurement error (classical or Berkson), missing data, or both are present in continuous covariates. Model fitting is done in a Bayesian framework using integrated nested Laplace approximations (INLA), an approach that is growing in popularity due to its combination of computational speed and accuracy. The \pkg{inlamemi} \proglang{R} package is suitable for data analysts who have little prior experience using the \proglang{R} package \pkg{R-INLA}, and aids in formulating suitable hierarchical models for a variety of scenarios in order to appropriately capture the processes that generate the measurement error and/or missingness. Numerous examples are given to help analysts identify scenarios similar to their own, and make the process of specifying a suitable model easier.
}

\Plainkeywords{missing data, classical measurement error, Berkson measurement error, R-INLA, Bayesian modelling}

\Address{
  Emma Skarstein\\
  Department of Mathematical Sciences\\
  Norwegian University of Science and Technology\\ 
  Norway\\
  E-mail: \email{emma@skarstein.no}\\
  URL: \url{https://emmaskarstein.github.io/}
}
\IfFileExists{upquote.sty}{\usepackage{upquote}}{}
\begin{document}



\section{Introduction} \label{sec:intro}

Measurement error and missing observations both mask the true value of our data, but exactly how they influence statistical inference may not always be obvious. A common claim is that the absolute value of a regression coefficient for a variable that is measured with error will be underestimated. Although such an attenuation effect is often observed, certain aspects like correlations between covariates, interaction effects, or the type of regression model may determine the direction of the bias, which makes it impossible to provide general rules that are true for all measurement error scenarios. Biased or inaccurate estimates are also a risk when using data with missing observations, along with reduced statistical power as a result of a smaller sample size.

Since measurement error arises in a wide range of situations, and error-generating mechanisms can be very diverse, the topic has been studied in many different contexts, and there exists a large body of work on the topic. 
After the seminal book by \citet{fuller1987}, \citet{carroll_etal2006} gave a more modern and comprehensive introduction to the problems different types of measurement error causes and how these can be modelled, but there are also some more recent books \citep{buonaccorsi2010, yi2017}. Shorter overviews over measurement error problems and methods are given by, for example, \citet{loken_gelman2017}, \citet{vansmeden_etal2019} or \citet{innes_etal2021}.

On the topic of missing data, on the other hand, there exists a separate body of literature. The classic book by \citet{little_rubin1987}, for example, provides a thorough guide, while some more recent books also give approachable advice, such as \citet{vanbuuren2018}. For a shorter read, summarising articles on the topic of missing data are \citet{nakagawa_freckleton2008} or \citet{graham2009}. 
Of particular note is the recent STRATOS initiative, which has established working groups covering (among other topics) measurement error and missing data seperately. These groups both have publications on their respective theme that summarize the fields quite concisely, see \citet{lee_etal2021} for missing data and \citet{stratos1_2020} for measurement error. 

When dealing with missing data we consider three main missingness mechanisms \citep{little_rubin1987}. A missingness mechanism can be described as a logistic regression model for the probability that a given observation is missing, and this probability can depend on other variables. The first mechanism is referred to as missing completely at random (MCAR), which means that the probability of an observation missing is independent of any other variables. Second, observations are missing at random (MAR) if the missingness probability depends on other observed variables. And third, observations are missing not at random (MNAR) if the probability of them missing depends on the value of the variable that is itself missing, for instance if certain individuals in a survey are not reporting their age, and the probability of a person not reporting their age increases with their age. 

Conceptually, we can think of missing data as an extreme case of a certain type of measurement error \citep{blackwell_etal2017}. On one end of the spectrum, a precisely observed variable would correspond to zero measurement error. As the measurement error increases, the correct value of the variable becomes more and more obscured, eventually leading to virtually no information about the correct value of the measured variable, which is analogous to the case of missing data. Taking advantage of this connection, we can model measurement error and missing data in the same framework, as pointed out by , for example, \citet{blackwell_etal2017} or \citet{skarstein_etal2023}.

Bayesian hierarchical models are particularly suitable for modelling measurement error and missingness (either separately  or jointly), because they can model the actual process generating the error or missingness mechanism, which then adjusts for eventual bias caused by it. In a Bayesian context, the missing values are viewed as further parameters to be estimated \citep[\eg][]{mcelreath2016}. Bayesian measurement error modelling was first detailed in \citet{stephens_dellaportas1992} and \citet{richardson_gilks1993A}. \citet{bartlett_keogh2018} compare frequentist and Bayesian methods for measurement error correction, and conclude that Bayesian measurement error models have a number of practical advantages such as being very flexible, allowing for a wide variety of complex models, and being fairly easy to implement.  Meanwhile, \citet{mason_etal2012}, \citet{erler_etal2016} and \citet{gomezrubio_etal2022} cover missing data in a Bayesian setting. Additionally, some more recent publications account for both measurement error and missing data, notably \citet{blackwell_etal2017}, \citet{goldstein_etal2018} and \citet{keogh_bartlett2019}. 

Integrated nested Laplace approximations (INLA) form the basis of a framework for approximate Bayesian inference that has been growing fast in popularity in recent years \citep{rue_etal2009, rue_etal2017, martino_riebler2020}. The main advantage with using INLA is the vast improvement in speed for certain models over traditional Markov chain Monte Carlo (MCMC) sampling based approaches. INLA is more restricted than MCMC methods in the class of models it can fit, because it requires that the model is a latent Gaussian model \citep[LGM,][]{rue_etal2009}. LGMs include a wide variety of standard  models, such as generalised linear mixed models (GLMMs), survival models, spatial and temporal models, and many more \citep{rue_held2005}. There are also some practical limitations to \pkg{R-INLA} that become apparent for models in which one of the covariates is latent, which is the case for measurement error and missing data, namely because the correct value for the covariate has not been observed but will instead be described by another level in the model. Such models are possible to fit using \pkg{R-INLA}, as described by \citet{muff_etal2015} and \citet{skarstein_etal2023}, but the implementation is not very straightforward and requires a good understanding of the inner workings of \pkg{R-INLA}. Models accounting for measurement error and missing data in INLA are therefore unfortunately less accessible than standard statistical models. 

As of now, there exists one other package that does missing data imputation with INLA; \pkg{MIINLA} \citep{gomezrubio_etal2022}. In this package, the variable with missingness is defined as a latent effect with a Gaussian Markov random field \citep[GMRF,][]{rue_held2005} structure. The authors of the package also show how to include a model describing the missingness mechanism, that is, a model that describes the probability of a given observation missing or not, which is modelled jointly with the imputation model and main regression model of interest. Additionally, approaches using INLA in combination with MCMC \citep{gomezrubio_rue2018} or importance sampling \citep{berild_etal2022} have also been proposed, but these quickly become computationally prohibitive as the number of missing observation grows. For measurement error alone, some simple models are implemented in \pkg{R-INLA} \citep{muff_etal2015}, but these are not very flexible in the way they allow to model the true distribution of the mis-observed covariate depending on any other variables we may have, which is commonly referred to as the exposure model.


The \pkg{inlamemi} \proglang{R} package aims to overcome the limitations of these existing solutions by providing an interface that is easy to use, and which works for a wider variety of error types. Our package provides functionality to account for classical measurement error, Berkson measurement error and missing observations (or any combination of these) in the covariates for many models that can be fit using \pkg{R-INLA}, specifically generalised linear mixed models and certain survival models. \pkg{inlamemi} consists of functions that automatically structure the data and set up the necessary additional components in the respective joint model formulation in the syntax needed by \pkg{R-INLA}, so that the user does not need to explicitly specify those parts of the model. The package can also easily incorporate repeated measurements of the variable with error, if those are available. If one has reason to believe that multiple independent variables have measurement error or missingness, this can also be accounted for. The structure of the models used in \pkg{inlamemi} are described in further detail in \citet{skarstein_etal2023}. Since each measurement error problem is unique, it is not possible to provide a universal tool that covers all scenarios. However, together with the package we provide a number of examples, as well as some illustrations of how to evaluate the sensitivity of the measurement error or missing data model to different assumptions, and how to evaluate if the errors actually do influence the conclusions of the analysis in a way that needs to be accounted for.

In the remainder of this paper, we describe the general model that \pkg{inlamemi} can fit, with emphasis on some common cases. We then give a description of how the package works more technically, describing the main function that fits the model, and some additional functionality. Finally, we provide two worked examples, one for a scenario with measurement error, and one with missing observations. 



\section{A model for missing data and measurement error} \label{sec:model}

\subsection{Model description}

We first look at the model of interest (MOI) that we would like to fit if there was no measurement error or missing observations. We assume that the linear predictor $g(\bm{\mu}) = \bm{\eta}$ in a generalised linear model (GLM) depends on a covariate $\bm{x}$ and a set of other covariates $\bm{Z}$ according to the model
\begin{equation}
  \bm{\eta} = \beta_0 \bm{1} + \beta_x \bm{x} + \bm{Z} \bm{\beta}_z \ ,  \label{eq:regmod}
\end{equation}

%
where $\bm{x}$ cannot be observed precisely, while the covariates $\bm{Z}$ can. Instead of $\bm{x}$, we observe a version of $\bm{x}$ that we call $\bm{w}$, which is connected to $\bm{x}$ through some error-generating mechanism. In the most simple case, the respective mechanism might be represented by a \emph{classical measurement error} model, where $\bm{x}$ is observed with independent additive noise such that $\bm{w}$ is given by
\begin{equation}
  \bm{w} = \bm{x} + \bm{u_c}  \ .  \label{eq:cmod}
\end{equation}
Here, $\bm{u_c} \sim \mathcal{N}(\bm{0}, \tau_{u_c}\bm{D}_{u_c})$ is an additive error term with a zero-mean Gaussian distribution and error precision $\tau_{u_c}\bm{D}_{u_c}$, where the matrix $\bm{D}_{u_c}$ may just be the identity matrix, meaning that all observations have the same error precision. Alternatively, the elements of $\bm{D}_{u_c}$ may encode heteroscedastic measurement error on a diagonal matrix with observation-dependent entries, or even a correlation structure between the entries of 
$\bm{x}$. 

When accounting for measurement error, we also include a third level in addition to the model of interest and the error model, which is typically referred to as the \emph{exposure model} in the measurement error literature. The exposure model describes how the correct value of the misobserved variable may possibly depend on other observed covariates, so we have a model for $\bm{x}$ through
\begin{equation}
\bm{x} = \alpha_0 \bm{1} + \widetilde{\bm{Z}}\bm{\alpha}_z + \bm{\varepsilon}_x \,, \qquad \bm{\varepsilon}_x \sim \mathcal{N}(\bm{0}, \tau_{\varepsilon_x}\bm{D}_{\varepsilon_x}) \ , \label{eq:expmod}
\end{equation}
where $\widetilde{\bm{Z}}$ is a matrix of covariates that are assumed to be observed without error, and which may or may not be the same as the covariates $\bm{Z}$ used in the model of interest, and $\tau_{\varepsilon_x}\bm{D}_{\varepsilon_x}$ is the precision for $\bm{\varepsilon}_x$. Again, the matrix $\bm{D}_{\varepsilon_x}$ may be the identity matrix, diagonal or encode for correlations among observations.
In the \pkg{inlamemi} package (both in the documentation and argument names) we will be referring to the coefficients of the model of interest as $\bm{\beta}$ and the coefficients of the exposure model as $\bm{\alpha}$. 

Another, fundamentally different, error mechanism is reflected by the \emph{Berkson measurement error} \citep{berkson1950}. In contrast to the classical error model, the Berkson measurement error model assumes that the recorded values actually have \emph{less} variance than the set of correct values, for instance because a shared treatment value has been recorded for multiple observations, when in reality the treatments are more variable than intended (\eg \, the amount of fertilizer given to crops). In that case, the model describing the error mechanism is
\begin{equation}
  \bm{x} = \bm{w} + \bm{u_b} \,,  \label{eq:bmod}
\end{equation}
where $\bm{w}$ is independent of $\bm{u_b}$, and $\bm{u_b} \sim \mathcal{N}(\bm{0}, \tau_{u_b}\bm{D}_{u_b})$, giving the same flexibility in the error precision as for the classical error. Since the Berkson measurement error model directly describes $\bm{x}$, an additional exposure model is not required. In the case of Berkson measurement error, the joint model thus reduces to only two levels: the model of interest and the Berkson error model.
Other possible error mechanisms exist beyond these two, but we do not cover them here since they would further complicate the model fitting procedure to the point where it would be necessary to combine INLA with MCMC \citep{gomezrubio_rue2018} to tackle them appropriately. 

When it comes to the connection between measurement error and \emph{missing data}, a few previous publications have illustrated how missing data can be modelled together with measurement error \citep{blackwell_etal2017, goldstein_etal2018, keogh_bartlett2019}. The link between classical measurement error and missing data arises when the classical measurement error variance $1/\tau_{u_c}$ approaches infinity, leading to a scenario where we essentially have no information about the true value of the variable. In practice, once we have a classical measurement error model for a given covariate (as in Equation \eqref{eq:cmod}), any missing observations in that covariate can be imputed through the exposure model described in Equation \eqref{eq:expmod}. With that perspective, we can actually model missing data through the classical measurement error model in combination with the exposure model, also in cases where there is no measurement error. In the missing data scenario it is more natural to refer to the exposure model as an imputation model, since its function is to impute the missing values, so from here on and throughout the \pkg{inlamemi} package we refer to the exposure model only as the \emph{imputation model}.

While each of the three -- classical measurement error, Berkson measurement error and missing data -- may occur individually, it is not unlikely that two or more of them are present simultaneously in the same covariate. In that case, it is possible to expand the model to contain four levels: the model of interest, the Berkson error model, the classical error model, and the imputation model. The respective theory and models are described in detail in \citet{skarstein_etal2023}. 

In the presence of missing observations, it may be relevant to include an additional model level, if we have reason to believe that the probability of an observation missing is somehow correlated with the value of the variable with missingness. We refer to this model level as the \emph{missingness} model. Though this layer in the joint model was not discussed in \citet{skarstein_etal2023}, it is described in the context of INLA in \cite{gomezrubio_etal2022}. The missingness model has a binary response, where a $1$ indicates that the corresponding entry in the covariate is missing. Then the model can be described as %
\begin{align}
  \bm{m} &\sim \text{Bernoulli}(\bm{\pi}_m) \ , \\
  \text{logit}(\bm{\pi}_m) &= \gamma_0 + \bm{Z}_m \bm{\gamma}_Z + \bm{\gamma}_x \bm{x} \ ,
\end{align}
where $\bm{Z}_m$ is a matrix of any covariates that are observed without error and make sense to include in the model. In the cases where the missingness mechanism is either MCAR or MAR, the missingness mechanism is ignorable \citep[see \eg][Chapter 2.2.5]{vanbuuren2018}. In that case, it is not necessary to include the missingness model, since it does not affect the posterior model. However, since we usually cannot know what the exact missingness mechanism is, it can be useful to examine and compare different versions of the missingness model to see if different assumptions about the missingness greatly influence the results \citep{mason_etal2012}.

The models sketched out here describe the main framework that can be fit using \pkg{inlamemi}. In summary, the possible levels are a main model of interest, a classical measurement error model, a Berkson measurement error model, an imputation model and a missingness model, which can be combined to fit the given application. However, several straight-forward extensions are also possible to allow for more flexibility. \emph{Random effects} can be added to either the model of interest, the imputation model or the missingness model, using the same syntax as would be used in standard \pkg{R-INLA}. In the same way, \emph{interaction effects} can also be added to these model levels. If there are \emph{multiple variables with error or missingness} (or a combination), then there is no technical limitation to the number of error variables that can be included. However, it is important to note that every new variable with error or missingness introduces new model levels and hyperparameters. If the number of hyperparameters is too large, this can start impacting the speed of \pkg{R-INLA}, since these hyperparameters must be integrated over numerically. The exact number of hyperparameters that \pkg{R-INLA} can manage has increased since its beginning, but in 2016 it was said that the number should not exceed 20 \citep{rue_etal2017}.

%


\subsection{Other important aspects} 

There are some other important considerations that must be made when fitting Bayesian measurement error and missing data models. We outline them briefly here. 

\subsubsection{Identifiability and priors}
The measurement error models \eqref{eq:cmod} and \eqref{eq:bmod} are not \emph{identifiable}. In the case of classical measurement error (Equation \eqref{eq:cmod}), we are not able to decompose $\text{Var}(\bm{w}) = \text{Var}(\bm{x}) + \text{Var}(\bm{u}_c)$ in the absence of preliminary knowledge about $\text{Var}(\bm{u}_c)$, because the only variable we observe is $\bm{w}$. Analogously, for Berkson measurement error (Equation \eqref{eq:bmod}) we cannot straight away decompose $\text{Var}(\bm{x}) = \text{Var}(\bm{w}) + \text{Var}(\bm{u}_b)$ into the terms of its sum, since neither $\text{Var}(\bm{x})$ nor $\text{Var}(\bm{u}_b)$ are observed. A further discussion of identifiability in the measurement error context can be found in \citet{gustafson2005}. 

In order to ensure correct inference, additional information about the error variance must thus be provided to the model. In the presence of repeated measurements for some or all observations of a variable with classical measurement error, the variance between measurements for the same subject can be used to estimate $\text{Var}(\bm{u}_c)$. This case can be directly represented in a Bayesian hierarchical model, as will be illustrated in the example in Section \ref{ex:classicalME}. Alternatively, we may have some knowledge about the measurement error variance, for instance due to validation studies or expert knowledge. The respective knowledge should then be included in the model by providing an informative prior on the measurement error variance. 

\subsubsection{Imputation model choice}
The choice of the imputation model will likely affect the parameter estimates and should be considered carefully. Since the purpose is to impute the values for the missing or mismeasured observations, the covariates chosen for the imputation model should be variables that may plausibly correlate with the variable that has missingness or measurement error. It may be tempting to simply include all available variables, but as in any prediction scenario this may lead to overfitting and is not necessarily advisable.

\subsubsection{Sensitivity analysis and validation}
We very rarely know exactly what mechanism is causing the missingness in our data, and in particular we may not be able to tell if the missingnes is MNAR or not. Therefore, it may be relevant to preform a sensitivity analysis to evaluate how much the results change if changes are made to the priors, imputation model or missingness model. In the case where one suspects that the missingness may be MNAR, one may compare the results from a model that accounts for MNAR along with one that does not and evaluate whether or how the results change \citep[see][]{gomezrubio_etal2022}. \citet{mason_etal2012} describe a step-by-step structured approach to sensitivity analysis for missing data imputation in a Bayesian setting.

\section[How to use inlamemi]{How to use \pkg{inlamemi}} \label{sec:howto}

%

%
The package can be installed from GitHub by the following code:
\begin{knitrout}
\definecolor{shadecolor}{rgb}{0.969, 0.969, 0.969}\color{fgcolor}\begin{kframe}
\begin{alltt}
\hlcom{# install.packages("devtools")}
\hlstd{devtools}\hlopt{::}\hlkwd{install_github}\hlstd{(}\hlstr{"emmaSkarstein/inlamemi"}\hlstd{)}
\end{alltt}
\end{kframe}
\end{knitrout}

\subsection[inlamemi functions]{Model fitting function}
The \pkg{inlamemi} \proglang{R} package provides a main function, \code{fit_inlamemi()}, which wraps around \pkg{R-INLA}'s \code{inla()} function that actually fits the model. The function \code{fit_inlamemi()} takes many of the same arguments as \code{inla()}, but also some that are specific to the measurement error or missingness aspect of the model, for instance the error type (whether there is classical or Berkson measurement error, and whether or not there are missing observations), as well as formulas describing both the main model of interest and the imputation model. Arguments relating to the priors for the different measurement error-specific terms, such as the classical error precision, are also directly provided to this function. Table \ref{tab:fitinlamemi} provides a detailed explanation of all the arguments for \code{fit_inlamemi()}. Note that all Gaussian priors are parameterized by the precision (typically denoted by $\tau$) rather than variance (typically denoted by $\sigma^2$) in \pkg{R-INLA} and therefore also in \pkg{inlamemi}.

\begin{longtable}{p{5cm}p{9.2cm}}\toprule
Argument & Description \\ \midrule
\code{formula_moi} &  Formula for the main model of interest.\\[2mm]
\code{formula_imp} &  Formula for the imputation model. \\[2mm]
\code{formula_mis} &  Formula for the missingness model (optional). \\[2mm]
\code{data} &  The data frame containing all the variables used in the model. \\[2mm]
\code{error_type} & A string or vector of strings specifying the error type(s), which would be one or more of \code{"classical"}, \code{"berkson"} or \code{"missing"}. \\[2mm]
\code{error_variable} & A string specifying the variable with measurement error or missingness. \\[2mm]
\code{repeated_observations} & Are there repeated measurements of the variable with measurement error? If so, set to \code{TRUE}. The columns in the data set should be named by the name of the error variable, followed by a number ($1, 2, \dots$) indicating the repeat number. For instance \code{sbp1, sbp2} (see example in Section \ref{ex:classicalME}). \\[2mm]
\code{classical_error_scaling} & An optional scaling vector which scales the precision of the classical error term. This can be used in cases where the measurement error is heteroscedastic, meaning that the measurement error variance can be different for different observations. \\[2mm]
\code{prior.beta.error} & The prior for the coefficient of the error variable ($\beta_x$), specified as a 2-element vector of the mean and precision for a Gaussian distribution. \\[2mm]
\code{prior.gamma.error} & The prior for the coefficient of the error variable ($\gamma_x$) in the (optional) missingness model, specified as a 2-element vector of the mean and precision for a Gaussian distribution. \\[2mm]
\code{prior.prec....} & The parameters for the gamma priors for the precisions of each level of the model. These are only specified for the model levels that are present. \\[2mm]
\code{initial.prec....} & The initial values  for the precisions of each level of the model, that is, the values where the numerical optimization will start. As with the priors, these are only specified for the model levels that are present. \\[2mm]
\code{control.family....} & In some cases, it may be necessary to specify additional options for the precisions for each model level than just specifying \code{prior.prec....} as above. Examples include fitting a model of interest that is not Gaussian and requires additional parameters (for example a survival model), or fixing the measurement error precision to a given value. One may choose to only specify one model level through this option, and use \code{prior.prec....} and \code{initial.prec....} for the remaining levels. See vignettes "Survival data with repeated systolic blood pressure measurements", "Simulated examples" and "Influence of systolic blood pressure on coronary heart disease" for specific illustrations. \\[2mm]
\code{control.family} & The entire \code{control.family} argument, which is an argument in \code{inla()}, can be specified directly here, instead of using the other options for providing priors described above. This will be passed directly to the \code{inla()} function. \\[2mm]
\code{control.fixed} & This is where the priors for the coefficients of all the observed covariates are provided, in the same way as for \code{inla()}. If these are not provided, the default priors $\mathcal{N}(0, 10^{-4})$ are used. \\[2mm]
\code{...} & Any additional arguments to \code{inla()} can be passed if needed.\\[2mm] \bottomrule
\caption{Description of all arguments for the main function.}
\label{tab:fitinlamemi}
\end{longtable}

\subsubsection{Structuring the arguments when there are multiple error variables}
If there are multiple variables with measurement error or missingness, some of the arguments to \code{fit_inlamemi} will need to be specified as lists, where the order of the list elements indicate which error variable the respective arguments belong to. For instance, in a scenario where one variable has classical measurement error and missingness and another variable only has missingness, then the error type argument will be \code{error_type = list(c("classical", "missing"), "missing")}. Similarly, the formula for the imputation model (and optionally also the missingness model) will be a list of two formulas, and the arguments \code{repeated_observations}, \code{classical_error_scaling} and \code{prior.beta.error} will also be specified as lists. To provide the parameters for the precisions, the \code{control.family} argument will need to be specified manually. An example of how to do this can be found in the package vignette "Multiple variables with measurement error and missingness". 

\subsubsection[Class inlamemi and methods]{Class \pkg{inlamemi} and methods}
The function \code{fit_inlamemi()} returns an object with two classes: \code{inlamemi} and \code{inla}. By default, calling \code{summary()} or \code{plot()} produces an \code{inlamemi} summary or plot, respectively, which highlights the elements of the model that are unique to the measurement error or missing data setting. But the standard \code{inla} methods can also be used, if these are preferred by the user. If you would like to see the standard \pkg{R-INLA} summary of an \pkg{inlamemi} model, this can be done by the following:

\begin{knitrout}
\definecolor{shadecolor}{rgb}{0.969, 0.969, 0.969}\color{fgcolor}\begin{kframe}
\begin{alltt}
\hlkwd{class}\hlstd{(inlamemi_model)} \hlkwb{<-} \hlstr{"inla"}
\hlkwd{summary}\hlstd{(inlamemi_model)}
\end{alltt}
\end{kframe}
\end{knitrout}

The \pkg{inlamemi} \code{plot()} function produces a \pkg{ggplot2} object \citep{ggplot2}, which can then be modified with standard \pkg{ggplot2} functions. An illustration of how to make some basic modifications to the default plot object, for instance separating it into sub-plots for each model level, or converting the coefficient names to Greek letters, can be seen in the vignette "Modifying the default plot".

\section{Examples} \label{sec:examples}


In the following examples we illustrate how to apply the package to two specific scenarios: 1) a case with only classical measurement error, and 2) a simulated data set where some of the observations are missing, but no measurement error. Since different scenarios may require different considerations, it can be helpful to look at a wider selection of examples than what we are able to provide here. Therefore, the \pkg{inlamemi} package includes multiple vignettes that illustrate various other specific cases. The vignettes cover examples with combinations of classical and Berkson measurement error and missing data, as well as repeated measurements of the error-prone variable and different likelihood families. There are examples with linear regression, logistic regression, Poisson regression and a Weibull survival model. All examples can all be found on the package website \url{https://emmaskarstein.github.io/inlamemi}.

\subsection{Classical measurement error with repeated measurements}\label{ex:classicalME}

As a first example, we consider a data set from the Framingham heart study, where interest is centered around understanding the connection between a given set of measured variables and the presence of heart disease. The data set, originally presented in \citet{macmahon_etal1990}, consists of 641 male individuals, and for each individual we have two measurements of systolic blood pressure, as well as their smoking status. We fit a logistic regression model where the binary response encodes for whether or not (1/0) a patient has heart disease, using systolic blood pressure (SBP) and smoking status as predictor variables. SBP is measured with error, and since we have two measurements per individual, we would like to use both repeats in the model. The data used in this example is also included in the package, and more details can be found through the command \code{?framingham}. The example has also been used in \citet{muff_etal2015}, and further details on how the model is directly encoded in \pkg{R-INLA} can be found there.

\subsubsection{Model description}\label{ex1_model}

We model disease status $y_i$ for individual $i$ with logistic regression as
\begin{align}
  y_i &\sim \text{Bernoulli}(\pi_{\texttt{disease}, i}) \ , \\
  \text{logit}\{\pi_{\texttt{disease}, i}\} &= \beta_0 + \beta_{\texttt{sbp}} \texttt{sbp}_i + \beta_{\texttt{smoking}} \texttt{smoking}_i \ ,
\end{align}
where $\pi_{\texttt{disease}, i}$ is the probability of individual $i$ having heart disease.
The classical measurement error model, which describes the actual error in the SBP measurements, has two levels in order to capture both repeats
\begin{align}
  \texttt{sbp}^1_i = \texttt{sbp}_i + u_i^{1} \ , \\
  \texttt{sbp}^2_i = \texttt{sbp}_i + u_i^{2} \ , 
\end{align}
with independent measurment error terms $u_i^{1}, u_i^{2} \sim N(0, \tau_u)$. Lastly, the imputation model is given as
\begin{equation}
  \texttt{sbp}_i = \alpha_0 + \alpha_{\texttt{smoking}} \texttt{smoking}_i + \varepsilon_i^{\texttt{sbp}} \ ,
\end{equation}
with error term $\varepsilon_i^{\texttt{sbp}} \sim \mathcal{N}(0, \tau_{\texttt{sbp}})$, where $\tau_{\texttt{sbp}}$ is the precision of the error term.

\subsubsection{Fitting the model}

We can call the \code{fit_inlamemi()} function directly with the above formulas for the model of interest and imputation model. The formulas are provided to the function just as they would be to other model fitting functions such as \code{lm()} or \code{inla()}. Note also that the \code{repeated_observations} argument, which is \code{FALSE} by default, must be set to \code{TRUE} to ensure that the model correctly captures the two repeats for the error-prone SBP variable. 

We give the precisions for the error and imputation models gamma priors. The precision for the error term of the measurement error model is given a $\tau_u\sim\text{G}(100, 1)$ prior, and the precision of the error term for the imputation model $\tau_{\texttt{sbp}} \sim \text{G}(10, 1)$. See \citet{muff_etal2015} for details on how the priors for these precision parameters are chosen. For $\beta_0$ and $\beta_{\texttt{smoking}}$, we assign the default $\mathcal{N}(0, 0.001)$ priors, and for $\beta_{\texttt{sbp}}$ we assign a $\mathcal{N}(0, 0.01)$ prior. The coefficients in the imputation model are assigned the default $\alpha_0, \alpha_{\texttt{smoking}}\sim\mathcal{N}(0, 0.001)$ priors. The respective model is then coded as follows: 
\begin{knitrout}
\definecolor{shadecolor}{rgb}{0.969, 0.969, 0.969}\color{fgcolor}\begin{kframe}
\begin{alltt}
\hlstd{framingham_model} \hlkwb{<-} \hlkwd{fit_inlamemi}\hlstd{(}
  \hlkwc{formula_moi} \hlstd{= disease} \hlopt{~} \hlstd{sbp} \hlopt{+} \hlstd{smoking,}
  \hlkwc{formula_imp} \hlstd{= sbp} \hlopt{~} \hlstd{smoking,}
  \hlkwc{family_moi} \hlstd{=} \hlstr{"binomial"}\hlstd{,}
  \hlkwc{data} \hlstd{= framingham,}
  \hlkwc{error_type} \hlstd{=} \hlstr{"classical"}\hlstd{,}
  \hlkwc{repeated_observations} \hlstd{=} \hlnum{TRUE}\hlstd{,}
  \hlkwc{prior.beta.error} \hlstd{=} \hlkwd{c}\hlstd{(}\hlnum{0}\hlstd{,} \hlnum{0.01}\hlstd{),}
  \hlkwc{prior.prec.classical} \hlstd{=} \hlkwd{c}\hlstd{(}\hlnum{100}\hlstd{,} \hlnum{1}\hlstd{),}
  \hlkwc{prior.prec.imp} \hlstd{=} \hlkwd{c}\hlstd{(}\hlnum{10}\hlstd{,} \hlnum{1}\hlstd{),}
  \hlkwc{initial.prec.classical} \hlstd{=} \hlnum{100}\hlstd{,}
  \hlkwc{initial.prec.imp} \hlstd{=} \hlnum{10}\hlstd{)}
\end{alltt}
\end{kframe}
\end{knitrout}

Once the model is fit we can print the and inspect the summary:
\begin{knitrout}
\definecolor{shadecolor}{rgb}{0.969, 0.969, 0.969}\color{fgcolor}\begin{kframe}
\begin{alltt}
\hlkwd{summary}\hlstd{(framingham_model)}
\end{alltt}
\end{kframe}
\end{knitrout}
\begin{knitrout}
\definecolor{shadecolor}{rgb}{0.969, 0.969, 0.969}\color{fgcolor}\begin{kframe}
\begin{verbatim}
## Formula for model of interest: 
## disease ~ sbp + smoking
## 
## Formula for imputation model: 
## sbp ~ smoking
## 
## Error types: 
## [1] "classical"
## 
## Fixed effects for model of interest: 
##                  mean      sd 0.025quant 0.5quant 0.975quant     mode
## beta.0       -2.36261 0.26912   -2.89195 -2.36203   -1.83652 -2.36202
## beta.smoking  0.40087 0.29801   -0.18311  0.40073    0.98569  0.40072
## 
## Coefficient for variable with measurement error and/or missingness: 
##           mean      sd 0.025quant 0.5quant 0.975quant   mode
## beta.sbp 1.905 0.56195    0.84752   1.8892     3.0579 1.8182
## 
## Fixed effects for imputation model: 
##                        mean       sd 0.025quant  0.5quant 0.975quant
## alpha.sbp.0        0.014545 0.018581  -0.021903  0.014545   0.050992
## alpha.sbp.smoking -0.019587 0.021563  -0.061882 -0.019587   0.022709
##                        mode
## alpha.sbp.0        0.014545
## alpha.sbp.smoking -0.019587
## 
## Model hyperparameters (apart from beta.sbp): 
##                                     mean     sd 0.025quant 0.5quant
## Precision for sbp classical model 75.902 3.6905     68.893   75.813
## Precision for sbp imp model       19.895 1.2340     17.551   19.865
##                                   0.975quant   mode
## Precision for sbp classical model     83.420 75.639
## Precision for sbp imp model           22.407 19.825
\end{verbatim}
\end{kframe}
\end{knitrout}

The summary contains an overview of the formulas and error type provided. The summary of the posterior marginals is structured into the fixed effects for the model of interest (here $\beta_0$ and $\beta_{\texttt{smoking}}$), the effect of the covariate with error (here $\beta_{\texttt{sbp}}$) and the fixed effects for the imputation model (here $\alpha_0$ and $\alpha_{\texttt{smoking}}$). At the end, summary statistics for the posterior marginals of the precisions for all model levels (here $\tau_u$ and $\tau_{\text{sbp}}$) are included in the chunk that contains the model hyperparameters. Note that, internally, the coefficient for the error prone variable ($\beta_{\texttt{sbp}}$) is also a hyperparameter, but for the sake of clarity it is printed in its own section in the summary.

\subsubsection[Comparing to a naive model and an MCMC approach using NIMBLE]{Comparing to a naive model and an MCMC approach using \pkg{NIMBLE}}

For comparison, and in order to  examine whether the measurement error affects the estimates from the model, we can also fit a so-called \emph{naive model}, that is, a model that ignores the measurement error in SBP. For the naive model, we use the average of the two SBP measurements per individual as a proxy for the SBP.
\begin{knitrout}
\definecolor{shadecolor}{rgb}{0.969, 0.969, 0.969}\color{fgcolor}\begin{kframe}
\begin{alltt}
\hlstd{framingham}\hlopt{$}\hlstd{sbp} \hlkwb{<-} \hlstd{(framingham}\hlopt{$}\hlstd{sbp1} \hlopt{+} \hlstd{framingham}\hlopt{$}\hlstd{sbp2)}\hlopt{/}\hlnum{2}
\end{alltt}
\end{kframe}
\end{knitrout}
The model is then fit using the standard \pkg{R-INLA} function \code{inla}, using the default priors for all coefficients:
\begin{knitrout}
\definecolor{shadecolor}{rgb}{0.969, 0.969, 0.969}\color{fgcolor}\begin{kframe}
\begin{alltt}
\hlstd{naive_model} \hlkwb{<-} \hlkwd{inla}\hlstd{(}\hlkwc{formula} \hlstd{= disease} \hlopt{~} \hlstd{sbp} \hlopt{+} \hlstd{smoking,}
                    \hlkwc{family} \hlstd{=} \hlstr{"binomial"}\hlstd{,}
                    \hlkwc{data} \hlstd{= framingham)}
\end{alltt}
\end{kframe}
\end{knitrout}

Additionally, we fit the same measurement error model as the one described in Section \ref{ex1_model}, but instead of \pkg{inlamemi} using \pkg{NIMBLE} \citep{nimble} to generate MCMC samples. We run one chain of 25000 iterations, and discard the first 5000 iterations as burn-in.
Both \pkg{inlamemi} and NIMBLE produce very similar point estimates, which show a slight adjustment for the coefficient of the error-prone variable, $\beta_{\texttt{sbp}}$ (Figure \ref{framingham_comparison}). However, while \pkg{inlamemi} took 3.84 seconds to run, \pkg{NIMBLE} took 63.71 seconds for the setup used here.

\begin{knitrout}
\definecolor{shadecolor}{rgb}{0.969, 0.969, 0.969}\color{fgcolor}\begin{figure}

{\centering \includegraphics[width=\maxwidth]{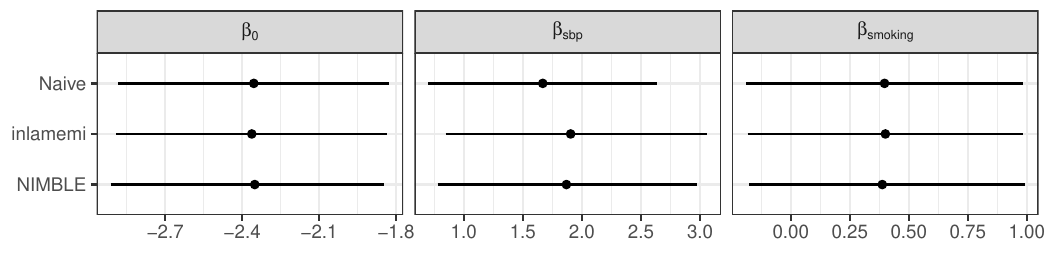} 

}

\caption[\label{framingham_comparison} Posterior means and 95\% credible intervals for the Framingham data analysis, comparing a model without any error adjustment ('Naive') to a model that adjusts for the measurement error, fit in \pkg{inlamemi} and \pkg{NIMBLE}]{\label{framingham_comparison} Posterior means and 95\% credible intervals for the Framingham data analysis, comparing a model without any error adjustment ('Naive') to a model that adjusts for the measurement error, fit in \pkg{inlamemi} and \pkg{NIMBLE}.}\label{fig:framingham_comparison}
\end{figure}

\end{knitrout}

\subsection{Missing data without measurement error}\label{sec:missing}

In this example, we simulate a simple data set with missingness in one of the covariates. 

\subsubsection{Model description}

The data set consists of the response variable $\bm{y}$, as well as three covariates, $\bm{x}$, $\bm{z}_1$ and $\bm{z}_2$. The variable $\bm{x}$ is missing in close to 20\% of the observations, and missingness is dependent on the value of $\bm{z}_2$, so the missingness mechanism is missing at random (MAR). The model for the response is
\begin{equation} \label{eq:ymis}
  \bm{y} = \beta_0 + \beta_x \bm{x} + \beta_{z_1} \bm{z}_1 + \beta_{z_2} \bm{z}_2 + \bm{\varepsilon} \ ,
\end{equation}
with $\beta_0 = 1$ and $\beta_x = \beta_{z_1} = \beta_{z_2} = 2$. 
To generate the data, we first set the seed and define the sample size.
\begin{knitrout}
\definecolor{shadecolor}{rgb}{0.969, 0.969, 0.969}\color{fgcolor}\begin{kframe}
\begin{alltt}
\hlkwd{set.seed}\hlstd{(}\hlnum{1}\hlstd{)}
\hlstd{n} \hlkwb{<-} \hlnum{1000}
\end{alltt}
\end{kframe}
\end{knitrout}

Next, we generate the covariates $\bm{z}_1$ and $\bm{z}_2$, which are not missing any observations. 
\begin{knitrout}
\definecolor{shadecolor}{rgb}{0.969, 0.969, 0.969}\color{fgcolor}\begin{kframe}
\begin{alltt}
\hlstd{z1} \hlkwb{<-} \hlkwd{rnorm}\hlstd{(n,} \hlkwc{mean} \hlstd{=} \hlnum{0}\hlstd{,} \hlkwc{sd} \hlstd{=} \hlnum{1}\hlstd{)}
\hlstd{z2} \hlkwb{<-} \hlkwd{rnorm}\hlstd{(n,} \hlkwc{mean} \hlstd{=} \hlnum{0}\hlstd{,} \hlkwc{sd} \hlstd{=} \hlnum{1}\hlstd{)}
\end{alltt}
\end{kframe}
\end{knitrout}

Once we have $\bm{z}_1$, we generate $\bm{x}$ according to the simple imputation model 

\begin{equation}
  \bm{x} = \alpha_0 + \alpha_{z_1}\bm{z}_1 + \bm{\varepsilon}_x \ ,
\end{equation}

where $\alpha_0 = 1$, $\alpha_{z_1} = 0.3$ and $\bm{\varepsilon}_x \sim \mathcal{N}(\bm{0}, \bm{I})$.
\begin{knitrout}
\definecolor{shadecolor}{rgb}{0.969, 0.969, 0.969}\color{fgcolor}\begin{kframe}
\begin{alltt}
\hlstd{x} \hlkwb{<-} \hlkwd{rnorm}\hlstd{(n,} \hlkwc{mean} \hlstd{=} \hlnum{1} \hlopt{+} \hlnum{0.3}\hlopt{*}\hlstd{z1,} \hlkwc{sd} \hlstd{=} \hlnum{1}\hlstd{)}
\end{alltt}
\end{kframe}
\end{knitrout}

We then remove some of the observations, where the missingness indicator $\bm{m}$ follows a Bernoulli distribution that depends on $\bm{z}_2$,
\begin{align}
  \bm{m} & \sim \text{Bernoulli}(\bm{\pi}_m) \ , \\
  \bm{\pi}_m & = \frac{\gamma_0 + \gamma_{z_2} \bm{z}_2}{1 + \gamma_0 + \gamma_{z_2} \bm{z}_2} \ ,
\end{align}
where $\gamma_0 = -1.5$ and $\gamma_{z_2} = -0.5$.

\begin{knitrout}
\definecolor{shadecolor}{rgb}{0.969, 0.969, 0.969}\color{fgcolor}\begin{kframe}
\begin{alltt}
\hlstd{m_pred} \hlkwb{<-} \hlopt{-}\hlnum{1.5} \hlopt{-} \hlnum{0.5}\hlopt{*}\hlstd{z2}
\hlstd{m_prob} \hlkwb{<-} \hlkwd{exp}\hlstd{(m_pred)}\hlopt{/}\hlstd{(}\hlnum{1} \hlopt{+} \hlkwd{exp}\hlstd{(m_pred))}
\hlstd{m} \hlkwb{<-} \hlkwd{as.logical}\hlstd{(}\hlkwd{rbinom}\hlstd{(n,} \hlnum{1}\hlstd{,} \hlkwc{prob} \hlstd{= m_prob))}
\hlstd{x_obs} \hlkwb{<-} \hlstd{x}
\hlstd{x_obs[m]} \hlkwb{<-} \hlnum{NA}
\end{alltt}
\end{kframe}
\end{knitrout}

This new version of $\bm{x}$, which we call $\bm{x}^\text{obs}$, is missing approximately 20\% of the observations:
\begin{knitrout}
\definecolor{shadecolor}{rgb}{0.969, 0.969, 0.969}\color{fgcolor}\begin{kframe}
\begin{alltt}
\hlkwd{sum}\hlstd{(}\hlkwd{is.na}\hlstd{(x_obs))}\hlopt{/}\hlkwd{length}\hlstd{(x_obs)}
\end{alltt}
\begin{verbatim}
## [1] 0.213
\end{verbatim}
\end{kframe}
\end{knitrout}

Finally, we generate the response $\bm{y}$, which depends on the unobserved variable $\bm{x}$ without missing observations as described in the model of interest in Equation \eqref{eq:ymis}.
\begin{knitrout}
\definecolor{shadecolor}{rgb}{0.969, 0.969, 0.969}\color{fgcolor}\begin{kframe}
\begin{alltt}
\hlstd{y} \hlkwb{<-} \hlnum{1} \hlopt{+} \hlnum{2}\hlopt{*}\hlstd{x} \hlopt{+} \hlnum{2}\hlopt{*}\hlstd{z1} \hlopt{+} \hlnum{2}\hlopt{*}\hlstd{z2} \hlopt{+} \hlkwd{rnorm}\hlstd{(n,} \hlkwc{mean} \hlstd{=} \hlnum{0}\hlstd{,} \hlkwc{sd} \hlstd{=} \hlnum{1}\hlstd{)}
\end{alltt}
\end{kframe}
\end{knitrout}

We save all the variables in a data frame.
\begin{knitrout}
\definecolor{shadecolor}{rgb}{0.969, 0.969, 0.969}\color{fgcolor}\begin{kframe}
\begin{alltt}
\hlstd{missing_data} \hlkwb{<-} \hlkwd{data.frame}\hlstd{(}\hlkwc{y} \hlstd{= y,} \hlkwc{x} \hlstd{= x_obs,} \hlkwc{x_true} \hlstd{= x,} \hlkwc{z1} \hlstd{= z1,} \hlkwc{z2} \hlstd{= z2)}
\end{alltt}
\end{kframe}
\end{knitrout}

\subsubsection{Fitting the model}

As before, we use the function \code{fit_inlamemi()} to fit the model. Here, we simulated $\bm{x}$ such that the value of $\bm{x}$ depends on $\bm{z}_1$ and the missingness probability depends on $\bm{z}_2$, but this is information we would not have if we had not simulated the data, so we use both available covariates in both the imputation model and the missingness model. Since we have no measurement error we only specify \code{error_type = "missing"}. Behind the scenes, this means that for the observations that are not missing, we scale the precision of the classical measurement error to be quite large, indicating that those observations do not have any measurement error \citep[for details, see][]{skarstein_etal2023}.

\begin{knitrout}
\definecolor{shadecolor}{rgb}{0.969, 0.969, 0.969}\color{fgcolor}\begin{kframe}
\begin{alltt}
\hlstd{missing_model} \hlkwb{<-} \hlkwd{fit_inlamemi}\hlstd{(}\hlkwc{formula_moi} \hlstd{= y} \hlopt{~} \hlstd{x} \hlopt{+} \hlstd{z1} \hlopt{+} \hlstd{z2,}
                              \hlkwc{formula_imp} \hlstd{= x} \hlopt{~} \hlstd{z1} \hlopt{+} \hlstd{z2,}
                              \hlkwc{formula_mis} \hlstd{= m} \hlopt{~} \hlstd{z1} \hlopt{+} \hlstd{z2} \hlopt{+} \hlstd{x,}
                              \hlkwc{family_moi} \hlstd{=} \hlstr{"gaussian"}\hlstd{,}
                              \hlkwc{data} \hlstd{= missing_data,}
                              \hlkwc{error_type} \hlstd{=} \hlstr{"missing"}\hlstd{,}
                              \hlkwc{prior.beta.error} \hlstd{=} \hlkwd{c}\hlstd{(}\hlnum{0}\hlstd{,} \hlnum{0.001}\hlstd{),}
                              \hlkwc{prior.gamma.error} \hlstd{=} \hlkwd{c}\hlstd{(}\hlnum{0}\hlstd{,} \hlnum{0.001}\hlstd{),}
                              \hlkwc{prior.prec.moi} \hlstd{=} \hlkwd{c}\hlstd{(}\hlnum{0.01}\hlstd{,} \hlnum{0.01}\hlstd{),}
                              \hlkwc{prior.prec.imp} \hlstd{=} \hlkwd{c}\hlstd{(}\hlnum{1}\hlstd{,} \hlnum{0.00005}\hlstd{),}
                              \hlkwc{initial.prec.moi} \hlstd{=} \hlnum{4}\hlstd{,}
                              \hlkwc{initial.prec.imp} \hlstd{=} \hlnum{4}\hlstd{)}
\end{alltt}
\end{kframe}
\end{knitrout}

We can then look at the summary of the model:
\begin{knitrout}
\definecolor{shadecolor}{rgb}{0.969, 0.969, 0.969}\color{fgcolor}\begin{kframe}
\begin{alltt}
\hlkwd{summary}\hlstd{(missing_model)}
\end{alltt}
\end{kframe}
\end{knitrout}
\begin{knitrout}
\definecolor{shadecolor}{rgb}{0.969, 0.969, 0.969}\color{fgcolor}\begin{kframe}
\begin{verbatim}
## Formula for model of interest: 
## y ~ x + z1 + z2
## 
## Formula for imputation model: 
## x ~ z1 + z2
## 
## Formula for missingness model: 
## m ~ z1 + z2 + x
## 
## Error types: 
## [1] "missing"
## 
## Fixed effects for model of interest: 
##           mean      sd 0.025quant 0.5quant 0.975quant   mode
## beta.0  0.9501 0.04803     0.8558   0.9501      1.045 0.9501
## beta.z1 1.9669 0.03550     1.8973   1.9669      2.037 1.9669
## beta.z2 1.9756 0.03400     1.9089   1.9756      2.042 1.9756
## 
## Coefficient for variable with measurement error and/or missingness: 
##            mean      sd 0.025quant 0.5quant 0.975quant    mode
## beta.x  2.03678 0.03286     1.9721   2.0368     2.1015 2.03679
## gamma.x 0.04034 0.08420    -0.1253   0.0403     0.2062 0.04011
## 
## Fixed effects for imputation model: 
##               mean      sd 0.025quant 0.5quant 0.975quant    mode
## alpha.x.0  1.00682 0.03349     0.9412  1.00682    1.07250 1.00682
## alpha.x.z1 0.34655 0.03227     0.2833  0.34655    0.40985 0.34655
## alpha.x.z2 0.02708 0.03221    -0.0361  0.02708    0.09025 0.02708
## 
## Fixed effects for missingness model: 
##                mean      sd 0.025quant 0.5quant 0.975quant     mode
## gamma.x.0  -1.43266 0.11957   -1.67177 -1.43127    -1.2021 -1.43125
## gamma.x.z1  0.08989 0.08210   -0.07127  0.08994     0.2507  0.08994
## gamma.x.z2 -0.47980 0.07908   -0.63488 -0.47979    -0.3248 -0.47979
## 
## Model hyperparameters (apart from beta.x, gamma.x): 
##                                   mean      sd 0.025quant 0.5quant
## Precision for model of interest 1.0294 0.05142     0.9312   1.0284
## Precision for x classical model 1.1454 0.38273     0.6056   1.0770
## Precision for x imp model       0.9489 0.04396     0.8646   0.9482
##                                 0.975quant   mode
## Precision for model of interest      1.134 1.0268
## Precision for x classical model      2.091 0.9452
## Precision for x imp model            1.038 0.9473
\end{verbatim}
\end{kframe}
\end{knitrout}

We can use the \code{plot()} function to plot the 95\% credible intervals of the coefficients along with their posterior means, which produces the plot shown in Figure \ref{missing_model_plot}.
\begin{knitrout}
\definecolor{shadecolor}{rgb}{0.969, 0.969, 0.969}\color{fgcolor}\begin{kframe}
\begin{alltt}
\hlkwd{plot}\hlstd{(missing_model)}
\end{alltt}
\end{kframe}\begin{figure}

{\centering \includegraphics[width=\maxwidth]{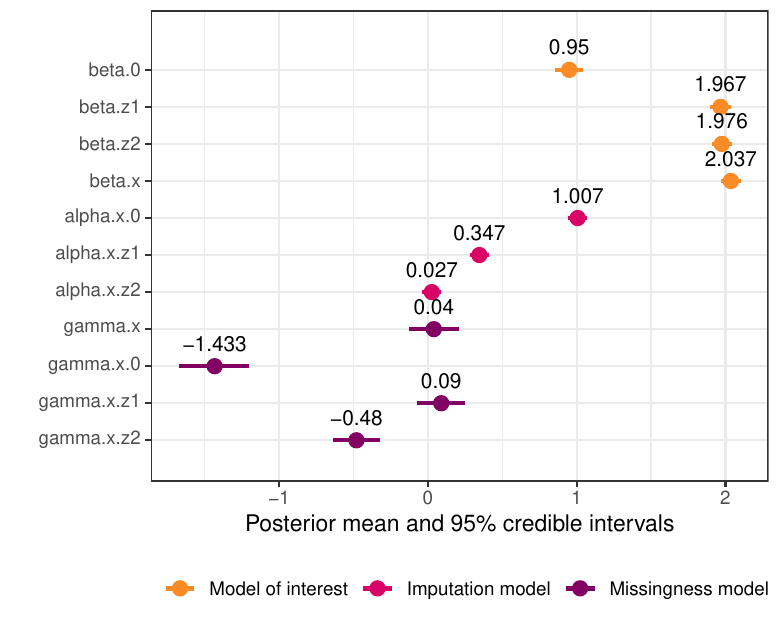} 

}

\caption[\label{missing_model_plot} The default plot, showing posterior means and 95\% credible intervals for the regression parameters in the different model levels of the missing data example]{\label{missing_model_plot} The default plot, showing posterior means and 95\% credible intervals for the regression parameters in the different model levels of the missing data example. The yellow credible intervals indicate the coefficients of the model of interest, the blue are for the imputation model, and the pink indicates the coefficient of the variable with missingness.}\label{fig:missing_model_plot}
\end{figure}

\end{knitrout}

We can also access the posterior marginal distributions of the imputed values by looking at \code{marginals.random$id.x} inside the model object (where \code{x} would be substituted by the name of the error variable).
The posterior marginal distributions for a few of the missing observations in the example are shown in Figure \ref{missing_values_plot}, along with a vertical line indicating the correct value for the missing observation from the simulation.

\begin{knitrout}
\definecolor{shadecolor}{rgb}{0.969, 0.969, 0.969}\color{fgcolor}\begin{figure}

{\centering \includegraphics[width=\maxwidth]{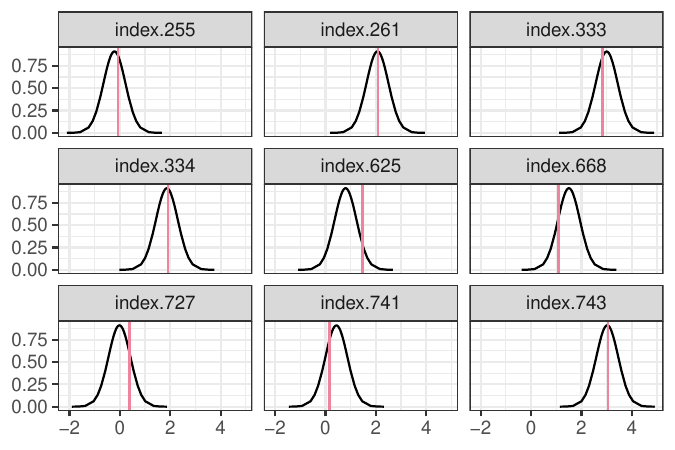} 

}

\caption{\label{missing_values_plot} Posterior marginal distributions for nine of the missing observations in Section \ref{sec:missing}, along with a vertical line indicating the correct value for each observation.}\label{fig:missing_values_plot}
\end{figure}

\end{knitrout}

\subsubsection{Comparison to other methods}

We also compare the missing data model to a complete case analysis fit in INLA, and to a model fit using the package \pkg{MIINLA} \citep{gomezrubio_etal2022}. As described in the introduction, \pkg{MIINLA} provides an alternative way for missing data imputation in \pkg{R-INLA}. \pkg{MIINLA} treats the variable with missing data as a latent effect with a GMRF structure. 
The joint distribution of the latent effect $\bm{x} = (\bm{x}_{mis}, \bm{x}_{obs})$ is defined by 
\begin{equation}
\bm{x} \mid \bm{\theta}_I = \mathcal{N}\left(
\begin{pmatrix}
\bm{\mu}_c \\
\bm{x}^{obs}
\end{pmatrix},
\begin{bmatrix}
\bm{Q}_c & \bm{0} \\
\bm{0} & 10^{10}\bm{I}
\end{bmatrix}
\right) \ ,
\end{equation}
where $\bm\theta_I$ is the set of hyperparameters for the imputation model and $\bm{x}^{obs}$ correspond to the elements of $\bm{w}$ (in the notation of this article) for which the value is not missing, while $\bm{x}^{mis}$ are the elements of $\bm{w}$ that are missing, as explained by \citet{gomezrubio_etal2022}. Further, $\bm{\mu}_c$ is the mean vector for the latent effects $\bm{x}^{mis}$, which can be suitably defined depending on the application, $\bm{Q}_c$ is the corresponding precision matrix for $\bm{x}_{mis}$, 
and $\bm{I}$ is an identity matrix with dimension equal to the number of observations that are not missing. In particular, in a linear regression model, the mean and precision of the elements in the latent effect corresponding to the missing values can be defined as
\begin{equation}
\bm{\mu}_c = \bm{Z}_{mis}\bm{\alpha} \ , \qquad \bm{Q}_c = \tau_x \bm{I}_{mis} \ ,
\end{equation}
where $\bm{Z}_{mis}$ is the matrix of covariates used in the imputation model, $\bm{\alpha}$ is the vector of corresponding coefficients, $\bm{I}_{mis}$ is an identity matrix with dimension equal to the number of missing values, and $\tau_x$ is the precision for the imputation model. Written like this, it is clear that the imputation model itself is the same as the one we adopt. The difference between \pkg{MIINLA} and the special case when \pkg{inlamemi} is used for only missing data lies in how they are implemented in \pkg{R-INLA}. 
In the case of \pkg{inlamemi}, the imputation model goes in as a layer in a hierarchical model, together with the model of interest and the error model. In \citet{gomezrubio_etal2022}, the imputation model is instead explicitly coded into the latent field through the \code{inla.rgeneric.define()} function. Details on this function can be found in, for example, Chapter 11 of \citet{gomezrubio2020}. 

Since we here used a simulated data set, we also fit a model using the correct observations without any missingness. The results for the different methods can be seen in Figure \ref{plot_missing_comparison}. By comparing the model using the correct value for $\bm{x}$ with the complete case model, we see that the missingness does not introduce any significant bias in $\bm{\beta}_x$, and the adjustments for both the \pkg{inlamemi} and \pkg{MIINLA} models are therefore minimal. More interestingly, we can also see that both the \pkg{inlamemi} and the \pkg{MIINLA} implementations pick up the MAR missingness mechanism, since both $\gamma_0$ and $\gamma_{z_2}$ are estimated very close to the correct values. Additionally, the $\gamma_x$ coefficient is estimated to be close to zero, which is as it should be, since we did not simulate the data to be MNAR. In this case, the differences between the methods are small and likely due to randomness inherent to \pkg{R-INLA}. In terms of speed, \pkg{inlamemi} is marginally faster, using 5.1 seconds, while \pkg{MIINLA} uses 11.6 seconds.

\begin{knitrout}
\definecolor{shadecolor}{rgb}{0.969, 0.969, 0.969}\color{fgcolor}\begin{figure}

{\centering \includegraphics[width=\maxwidth]{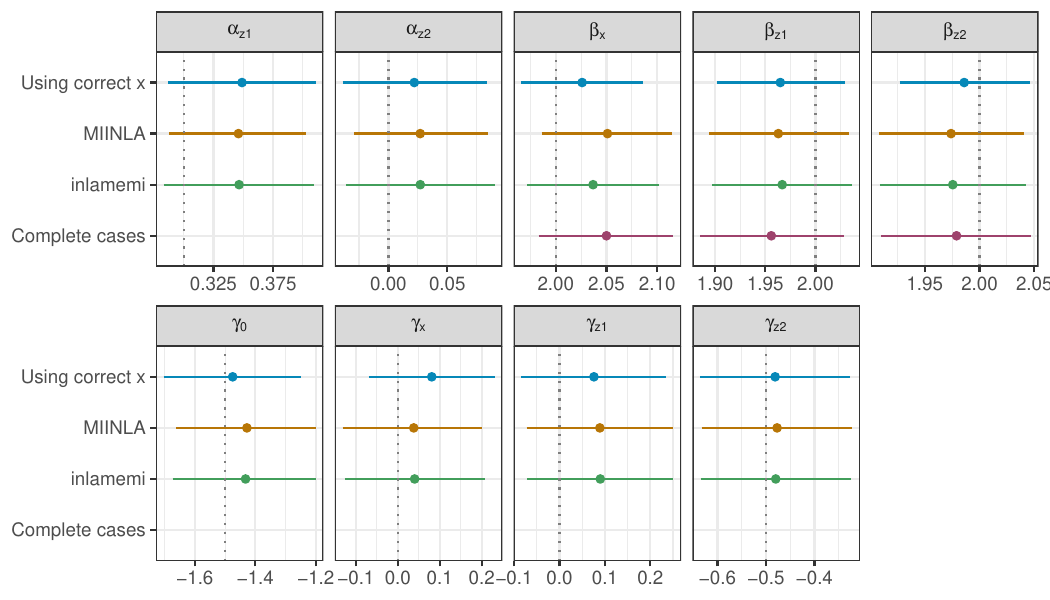} 

}

\caption[\label{plot_missing_comparison} Posterior means and 95\% credible intervals for the missing data analysis, comparing the results from fitting the same model in \pkg{inlamemi} and \pkg{MIINLA} to a model using only complete observations ('Complete cases') and one using the full data set without missingness ('Using correct x')]{\label{plot_missing_comparison} Posterior means and 95\% credible intervals for the missing data analysis, comparing the results from fitting the same model in \pkg{inlamemi} and \pkg{MIINLA} to a model using only complete observations ('Complete cases') and one using the full data set without missingness ('Using correct x'). The vertical dotted lines indicate the parameter values used in the simulation.}\label{fig:plot_missing_comparison}
\end{figure}

\end{knitrout}


\section{Summary and discussion} \label{sec:summary}

We present the R package \pkg{inlamemi}, which provides a wrapper function around the standard \pkg{R-INLA} model fitting function \code{inla()}, in order to assist formulating and fitting models that account for measurement error and missing data in covariates of regression models. The package implements the models presented in \citet{skarstein_etal2023}, with extensions to allow for modelling a missingness mechanism that is MNAR, in a convenient interface. The package currently works for GLMMs and some types of survival models with measurement error or missingness in continuous covariates. The package handles both classical and Berkson measurement error, and it is possible to account for variables that are subject to both error types and missingness at the same time, or multiple variables that may be affected by measurement error and/or missingness. It is also possible to include random effects or interaction effects in the predictors for the model of interest, the imputation model and the missingness model. The use of the package is illustrated on two relatively simple examples, but numerous additional examples are provided in the package vignettes, available online at \url{https://emmaskarstein.github.io/inlamemi/}.


In the comparison to the existing packages \pkg{NIMBLE} (which was used to fit a classical measurement error model for repeated measurements) and \pkg{MIINLA} (which was used to do missing data imputation), we found that \pkg{inlamemi} produces similar results as the other packages, but is faster to run. In addition, it 
caters to a greater variety of measurement error models and makes it very easy to account for multiple types of measurement error and missingness in the same variable. 
A current caveat of \pkg{inlamemi} is that the flexibility in the main model of interest is a bit smaller than, for example, for \pkg{MIINLA}. 
Some future extensions can be considered if this is of interest: \pkg{inlamemi} has not been tested with spatial or temporal models, but this would be entirely feasible to include due to the modular structure of \pkg{R-INLA}. Additionally, though the package currently only adjusts for measurement error and missingness in continuous covariates, it is possible to extend it to cover discrete values as well. See for instace the approach used in \citet{gomezrubio_etal2022} for details on modelling missingness in categorical covariates.

In summary, \pkg{inlamemi} provides fast and convenient implementations for a variety of measurement and missing data problems for continuous variables in a wide family of regression models. It is significantly faster than software that provides Bayesian implementations that rely on MCMC sampling, and it is at the same time accessible to users that are not so familiar with using \pkg{R-INLA}. For these reasons, \pkg{inlamemi} is a useful contribution for scientists looking to implement models for measurement error and missing data adjustment.


\section*{Computational details}

The results in this paper were obtained using \proglang{R}~4.3.1 
, \pkg{R-INLA} version 23.11.01 
and \pkg{inlamemi} version 1.0.0.


\bibliography{bibliography}


\newpage

\begin{appendix}

\section[Code for comparison in Section 4.1]{Code for comparison in Section 4.1} \label{app:comp_41}

In Section \ref{ex:classicalME}, we compare the model fit in \pkg{inlamemi} to the equivalent model fit in \pkg{NIMBLE}, but only the code needed to fit the \pkg{inlamemi} model is shown in the paper. In this appendix, we go through the \pkg{NIMBLE} code used in that comparison.

\begin{knitrout}
\definecolor{shadecolor}{rgb}{0.969, 0.969, 0.969}\color{fgcolor}\begin{kframe}
\begin{alltt}
\hlkwd{library}\hlstd{(nimble)}
\end{alltt}
\end{kframe}
\end{knitrout}

First, we need to define the model using the \pkg{NIMBLE} syntax:
\begin{knitrout}
\definecolor{shadecolor}{rgb}{0.969, 0.969, 0.969}\color{fgcolor}\begin{kframe}
\begin{alltt}
\hlstd{framingham_code} \hlkwb{<-} \hlkwd{nimbleCode}\hlstd{(\{}
  \hlkwa{for} \hlstd{(i} \hlkwa{in} \hlnum{1}\hlopt{:}\hlstd{n) \{}
    \hlcom{# Response model}
    \hlstd{disease[i]} \hlopt{~} \hlkwd{dbin}\hlstd{(pdisease[i],} \hlnum{1}\hlstd{)}
    \hlkwd{logit}\hlstd{(pdisease[i])} \hlkwb{<-} \hlstd{beta.0} \hlopt{+} \hlstd{beta.sbp}\hlopt{*}\hlstd{sbp[i]} \hlopt{+} \hlstd{beta.smoking}\hlopt{*}\hlstd{smoking[i]}

    \hlcom{# Error models}
    \hlstd{sbp1[i]} \hlopt{~} \hlkwd{dnorm}\hlstd{(sbp[i],} \hlkwc{tau} \hlstd{= tauu)}
    \hlstd{sbp2[i]} \hlopt{~} \hlkwd{dnorm}\hlstd{(sbp[i], tauu)}

    \hlcom{# Exposure model}
    \hlstd{sbp[i]} \hlopt{~} \hlkwd{dnorm}\hlstd{(alpha.0} \hlopt{+} \hlstd{alpha.smoking}\hlopt{*}\hlstd{smoking[i], tausbp)}
  \hlstd{\}}

  \hlcom{# Priors on the model parameters}
  \hlstd{tauu} \hlopt{~} \hlkwd{dgamma}\hlstd{(}\hlkwc{shape} \hlstd{=} \hlnum{100}\hlstd{,} \hlkwc{rate} \hlstd{=} \hlnum{1}\hlstd{)}
  \hlstd{tausbp} \hlopt{~} \hlkwd{dgamma}\hlstd{(}\hlnum{10}\hlstd{,} \hlnum{1}\hlstd{)}

  \hlstd{beta.0} \hlopt{~} \hlkwd{dnorm}\hlstd{(}\hlnum{0}\hlstd{,} \hlnum{0.001}\hlstd{)}
  \hlstd{beta.sbp} \hlopt{~} \hlkwd{dnorm}\hlstd{(}\hlnum{0}\hlstd{,} \hlnum{0.01}\hlstd{)}
  \hlstd{beta.smoking} \hlopt{~} \hlkwd{dnorm}\hlstd{(}\hlnum{0}\hlstd{,} \hlnum{0.001}\hlstd{)}

  \hlstd{alpha.0} \hlopt{~} \hlkwd{dnorm}\hlstd{(}\hlnum{0}\hlstd{,} \hlnum{0.0001}\hlstd{)}
  \hlstd{alpha.smoking} \hlopt{~} \hlkwd{dnorm}\hlstd{(}\hlnum{0}\hlstd{,} \hlnum{0.0001}\hlstd{)}
\hlstd{\})}
\end{alltt}
\end{kframe}
\end{knitrout}

Next, we need to define the data, constants and initial values, which are chosen to be the naive regression model estimates for the coefficients, and for the precisions are set to the same values as we used for the \pkg{inlamemi} model:
\begin{knitrout}
\definecolor{shadecolor}{rgb}{0.969, 0.969, 0.969}\color{fgcolor}\begin{kframe}
\begin{alltt}
\hlcom{# Data }
\hlstd{data} \hlkwb{<-} \hlkwd{list}\hlstd{(}\hlkwc{disease} \hlstd{= framingham}\hlopt{$}\hlstd{disease,}
             \hlkwc{sbp1} \hlstd{= framingham}\hlopt{$}\hlstd{sbp1,}
             \hlkwc{sbp2} \hlstd{= framingham}\hlopt{$}\hlstd{sbp2)}

\hlcom{# Constants}
\hlstd{const} \hlkwb{<-} \hlkwd{list}\hlstd{(}\hlkwc{n} \hlstd{=} \hlkwd{nrow}\hlstd{(framingham),}
              \hlkwc{smoking} \hlstd{= framingham}\hlopt{$}\hlstd{smoking)}

\hlcom{# Initial values}
\hlstd{inits} \hlkwb{<-} \hlkwd{list}\hlstd{(}\hlkwc{beta.0} \hlstd{=} \hlopt{-}\hlnum{2.4}\hlstd{,} \hlkwc{beta.sbp} \hlstd{=} \hlnum{1.67}\hlstd{,} \hlkwc{beta.smoking} \hlstd{=} \hlnum{0.40}\hlstd{,}
              \hlkwc{alpha.0} \hlstd{=} \hlnum{0}\hlstd{,} \hlkwc{alpha.smoking} \hlstd{=} \hlnum{0}\hlstd{,}
              \hlkwc{tausbp} \hlstd{=} \hlnum{10}\hlstd{,} \hlkwc{tauu} \hlstd{=} \hlnum{100}\hlstd{,} \hlkwc{sbp} \hlstd{= (data}\hlopt{$}\hlstd{sbp1}\hlopt{+}\hlstd{data}\hlopt{$}\hlstd{sbp2)}\hlopt{/}\hlnum{2}\hlstd{)}
\end{alltt}
\end{kframe}
\end{knitrout}

We can then fit the model:
\begin{knitrout}
\definecolor{shadecolor}{rgb}{0.969, 0.969, 0.969}\color{fgcolor}\begin{kframe}
\begin{alltt}
\hlstd{nimble_mod} \hlkwb{<-} \hlkwd{nimbleMCMC}\hlstd{(}\hlkwc{code} \hlstd{= framingham_code,}
                       \hlkwc{data} \hlstd{= data,}
                       \hlkwc{constants} \hlstd{= const,}
                       \hlkwc{inits} \hlstd{= inits,}
                       \hlkwc{niter} \hlstd{=} \hlnum{25000}\hlstd{,}
                       \hlkwc{nburnin} \hlstd{=} \hlnum{5000}\hlstd{,}
                       \hlkwc{summary} \hlstd{=} \hlnum{TRUE}\hlstd{,}
                       \hlkwc{monitors} \hlstd{=} \hlkwd{c}\hlstd{(}\hlstr{'beta.0'}\hlstd{,}\hlstr{'beta.sbp'}\hlstd{,}\hlstr{'beta.smoking'}\hlstd{,}
                                    \hlstr{'alpha.0'}\hlstd{,} \hlstr{'alpha.smoking'}\hlstd{))}
\end{alltt}
\end{kframe}
\end{knitrout}

And the summary can be viewed like this:
\begin{knitrout}
\definecolor{shadecolor}{rgb}{0.969, 0.969, 0.969}\color{fgcolor}\begin{kframe}
\begin{alltt}
\hlstd{nimble_mod}\hlopt{$}\hlstd{summary}
\end{alltt}
\end{kframe}
\end{knitrout}

\section[Code for comparison in Section 4.2]{Code for comparison in Section 4.2} \label{app:comp_42}

In Section \ref{sec:missing}, we simulated a data set and compared the \pkg{inlamemi} model to an implementation in the package \pkg{MIINLA}, along with a complete case model and a model using the correct version of the variable with missingness. The code below illustrates how each of the models were fit, apart from the \pkg{inlamemi} model, which was demonstrated in Section \ref{sec:missing}. 

\subsubsection{Complete case model with INLA}

For the complete case model, only the observations for which the corresponding entry in $\bm{x}$ is observed are used.
\begin{knitrout}
\definecolor{shadecolor}{rgb}{0.969, 0.969, 0.969}\color{fgcolor}\begin{kframe}
\begin{alltt}
\hlkwd{library}\hlstd{(INLA)}
\hlstd{cc_model} \hlkwb{<-} \hlkwd{inla}\hlstd{(y} \hlopt{~} \hlstd{z1} \hlopt{+} \hlstd{z2} \hlopt{+} \hlstd{x,}
                 \hlkwc{data} \hlstd{= missing_data[}\hlopt{!}\hlkwd{is.na}\hlstd{(missing_data}\hlopt{$}\hlstd{x), ])}
\end{alltt}
\end{kframe}
\end{knitrout}

\subsubsection[Missing data imputation with MIINLA]{Missing data imputation with \pkg{MIINLA}}
To fit the model with \pkg{MIINLA}, some structuring of the data is necessary.

\begin{knitrout}
\definecolor{shadecolor}{rgb}{0.969, 0.969, 0.969}\color{fgcolor}\begin{kframe}
\begin{alltt}
\hlkwd{library}\hlstd{(MIINLA)}

\hlstd{n} \hlkwb{<-} \hlkwd{nrow}\hlstd{(missing_data)}

\hlstd{missing_data}\hlopt{$}\hlstd{idxNA} \hlkwb{<-} \hlkwd{rep}\hlstd{(}\hlnum{NA}\hlstd{, n)}
\hlstd{missing_data}\hlopt{$}\hlstd{idx2} \hlkwb{<-} \hlnum{1}\hlopt{:}\hlstd{n}

\hlcom{# Response and missingness for x}
\hlstd{Y} \hlkwb{<-} \hlkwd{matrix}\hlstd{(}\hlnum{NA}\hlstd{,} \hlkwc{nrow} \hlstd{=} \hlnum{2} \hlopt{*} \hlstd{n,} \hlkwc{ncol} \hlstd{=} \hlnum{2}\hlstd{)}
\hlstd{Y[}\hlnum{1}\hlopt{:}\hlstd{n,} \hlnum{1}\hlstd{]} \hlkwb{<-} \hlstd{missing_data}\hlopt{$}\hlstd{y}
\hlstd{Y[n} \hlopt{+} \hlnum{1}\hlopt{:}\hlstd{n,} \hlnum{2}\hlstd{]} \hlkwb{<-} \hlkwd{is.na}\hlstd{(missing_data}\hlopt{$}\hlstd{x)}

\hlcom{# This will store x}
\hlstd{idxNA} \hlkwb{<-} \hlkwd{rep}\hlstd{(}\hlnum{NA}\hlstd{,} \hlnum{2} \hlopt{*} \hlstd{n)}
\hlcom{# This will store beta.x*x}
\hlstd{beta.x} \hlkwb{<-} \hlkwd{c}\hlstd{(}\hlnum{1}\hlopt{:}\hlstd{n,} \hlkwd{rep}\hlstd{(}\hlnum{NA}\hlstd{, n))}
\hlcom{# This will store gamma.x*x}
\hlstd{gamma.x} \hlkwb{<-} \hlkwd{c}\hlstd{(}\hlkwd{rep}\hlstd{(}\hlnum{NA}\hlstd{, n),} \hlnum{1}\hlopt{:}\hlstd{n)}

\hlcom{# Covarite for MOI and missingness model}
\hlstd{beta.0} \hlkwb{<-} \hlkwd{c}\hlstd{(}\hlkwd{rep}\hlstd{(}\hlnum{1}\hlstd{, n),} \hlkwd{rep}\hlstd{(}\hlnum{NA}\hlstd{, n))}
\hlstd{beta.z1} \hlkwb{<-} \hlkwd{c}\hlstd{(missing_data}\hlopt{$}\hlstd{z1,} \hlkwd{rep}\hlstd{(}\hlnum{NA}\hlstd{, n))}
\hlstd{beta.z2} \hlkwb{<-} \hlkwd{c}\hlstd{(missing_data}\hlopt{$}\hlstd{z2,} \hlkwd{rep}\hlstd{(}\hlnum{NA}\hlstd{, n))}

\hlstd{gamma.0} \hlkwb{<-} \hlkwd{c}\hlstd{(}\hlkwd{rep}\hlstd{(}\hlnum{NA}\hlstd{, n),} \hlkwd{rep}\hlstd{(}\hlnum{1}\hlstd{, n))}
\hlstd{gamma.z1} \hlkwb{<-} \hlkwd{c}\hlstd{(}\hlkwd{rep}\hlstd{(}\hlnum{NA}\hlstd{, n), missing_data}\hlopt{$}\hlstd{z1)}
\hlstd{gamma.z2} \hlkwb{<-} \hlkwd{c}\hlstd{(}\hlkwd{rep}\hlstd{(}\hlnum{NA}\hlstd{, n), missing_data}\hlopt{$}\hlstd{z2)}
\end{alltt}
\end{kframe}
\end{knitrout}

Once the data is structured, we can define the imputation model using \code{inla.rgeneric.define()}:

\begin{knitrout}
\definecolor{shadecolor}{rgb}{0.969, 0.969, 0.969}\color{fgcolor}\begin{kframe}
\begin{alltt}
\hlstd{imputation_model} \hlkwb{<-} \hlstd{INLA}\hlopt{::}\hlkwd{inla.rgeneric.define}\hlstd{(}
  \hlstd{inla.rgeneric.milm.model,}  \hlcom{# This comes from MIINLA}
  \hlkwc{debug} \hlstd{=} \hlnum{TRUE}\hlstd{,}
  \hlkwc{x} \hlstd{= missing_data}\hlopt{$}\hlstd{x,}
  \hlkwc{XX} \hlstd{=} \hlkwd{cbind}\hlstd{(}\hlnum{1}\hlstd{, missing_data}\hlopt{$}\hlstd{z1, missing_data}\hlopt{$}\hlstd{z2),}
  \hlkwc{n} \hlstd{=} \hlkwd{nrow}\hlstd{(missing_data),}
  \hlkwc{idx.na} \hlstd{=} \hlkwd{which}\hlstd{(}\hlkwd{is.na}\hlstd{(missing_data}\hlopt{$}\hlstd{x)))}
\end{alltt}
\end{kframe}
\end{knitrout}

The imputation model is then called inside the formula, as \code{f(idxNA, model = imputation_model)}:
\begin{knitrout}
\definecolor{shadecolor}{rgb}{0.969, 0.969, 0.969}\color{fgcolor}\begin{kframe}
\begin{alltt}
\hlstd{miinla.formula} \hlkwb{<-} \hlstd{Y} \hlopt{~ -}\hlnum{1} \hlopt{+} \hlstd{beta.0} \hlopt{+} \hlstd{beta.z1} \hlopt{+} \hlstd{beta.z2} \hlopt{+}
  \hlstd{gamma.0} \hlopt{+} \hlstd{gamma.z1} \hlopt{+} \hlstd{gamma.z2} \hlopt{+}
  \hlkwd{f}\hlstd{(idxNA,} \hlkwc{model} \hlstd{= imputation_model)} \hlopt{+}
  \hlkwd{f}\hlstd{(beta.x,} \hlkwc{copy} \hlstd{=} \hlstr{"idxNA"}\hlstd{,} \hlkwc{fixed} \hlstd{=} \hlnum{FALSE}\hlstd{,}
    \hlkwc{hyper} \hlstd{=} \hlkwd{list}\hlstd{(}\hlkwc{beta} \hlstd{=} \hlkwd{list}\hlstd{(}\hlkwc{prior} \hlstd{=} \hlstr{"normal"}\hlstd{,} \hlkwc{param} \hlstd{=} \hlkwd{c}\hlstd{(}\hlnum{0}\hlstd{,} \hlnum{0.001}\hlstd{))))} \hlopt{+}
  \hlkwd{f}\hlstd{(gamma.x,} \hlkwc{copy} \hlstd{=} \hlstr{"idxNA"}\hlstd{,} \hlkwc{fixed} \hlstd{=} \hlnum{FALSE}\hlstd{,}
    \hlkwc{hyper} \hlstd{=} \hlkwd{list}\hlstd{(}\hlkwc{beta} \hlstd{=} \hlkwd{list}\hlstd{(}\hlkwc{prior} \hlstd{=} \hlstr{"normal"}\hlstd{,} \hlkwc{param} \hlstd{=} \hlkwd{c}\hlstd{(}\hlnum{0}\hlstd{,} \hlnum{0.001}\hlstd{))))}
\end{alltt}
\end{kframe}
\end{knitrout}

Then the \code{inla()} function is called in order to fit the model. The two likelihoods passed to the \code{family} argument correspond to the model of interest and the missingness model.
\begin{knitrout}
\definecolor{shadecolor}{rgb}{0.969, 0.969, 0.969}\color{fgcolor}\begin{kframe}
\begin{alltt}
\hlstd{miinla.mod} \hlkwb{<-} \hlkwd{inla}\hlstd{(miinla.formula,} \hlkwc{family} \hlstd{=} \hlkwd{c}\hlstd{(}\hlstr{"gaussian"}\hlstd{,} \hlstr{"binomial"}\hlstd{),}
                  \hlkwc{verbose} \hlstd{=} \hlnum{FALSE}\hlstd{,}
                  \hlkwc{data} \hlstd{=} \hlkwd{list}\hlstd{(}\hlkwc{Y} \hlstd{= Y,}
                              \hlkwc{beta.0} \hlstd{= beta.0,} \hlkwc{beta.z1} \hlstd{= beta.z1,}
                              \hlkwc{beta.z2} \hlstd{= beta.z2,} \hlkwc{beta.x} \hlstd{= beta.x,}
                              \hlkwc{gamma.0} \hlstd{= gamma.0,} \hlkwc{gamma.z1} \hlstd{= gamma.z1,}
                              \hlkwc{gamma.z2} \hlstd{= gamma.z2,} \hlkwc{gamma.x} \hlstd{= gamma.x,}
                              \hlkwc{idxNA} \hlstd{= idxNA),}
                  \hlkwc{control.family} \hlstd{=} \hlkwd{list}\hlstd{(}
                    \hlkwd{list}\hlstd{(}\hlkwc{hyper} \hlstd{=} \hlkwd{list}\hlstd{(}\hlkwc{prec} \hlstd{=} \hlkwd{list}\hlstd{(}\hlkwc{param} \hlstd{=} \hlkwd{c}\hlstd{(}\hlnum{0.01}\hlstd{,} \hlnum{0.01}\hlstd{)))),}
                    \hlkwd{list}\hlstd{())}
\hlstd{)}
\end{alltt}
\end{kframe}
\end{knitrout}

When interpreting the summary, note that \code{Theta1} corresponds to $\alpha_0$, \code{Theta2} corresponds to $\alpha_{z_1}$, \code{Theta3} corresponds to $\alpha_{z2}$ and \code{Theta4} corresponds to $\text{log}(\tau_x)$. 
\begin{knitrout}
\definecolor{shadecolor}{rgb}{0.969, 0.969, 0.969}\color{fgcolor}\begin{kframe}
\begin{alltt}
\hlkwd{summary}\hlstd{(miinla.mod)}
\end{alltt}
\end{kframe}
\end{knitrout}

\subsubsection[Model using correct version of x]{Model using correct version of $\bm{x}$}

For the models using the correct version of $\bm{x}$, we fit the three layers separately in \pkg{R-INLA}:
\begin{knitrout}
\definecolor{shadecolor}{rgb}{0.969, 0.969, 0.969}\color{fgcolor}\begin{kframe}
\begin{alltt}
\hlstd{correct_moi} \hlkwb{<-} \hlkwd{inla}\hlstd{(y} \hlopt{~} \hlstd{z1} \hlopt{+} \hlstd{z2} \hlopt{+} \hlstd{x_true,} \hlkwc{data} \hlstd{= missing_data)}
\hlstd{correct_imp} \hlkwb{<-} \hlkwd{inla}\hlstd{(x_true} \hlopt{~} \hlstd{z1} \hlopt{+} \hlstd{z2,} \hlkwc{data} \hlstd{= missing_data)}
\hlstd{correct_mis} \hlkwb{<-} \hlkwd{inla}\hlstd{(}\hlkwd{as.numeric}\hlstd{(}\hlkwd{is.na}\hlstd{(x))} \hlopt{~} \hlstd{z1} \hlopt{+} \hlstd{z2} \hlopt{+} \hlstd{x_true,}
                    \hlkwc{data} \hlstd{= missing_data,} \hlkwc{family} \hlstd{=} \hlstr{"binomial"}\hlstd{)}
\end{alltt}
\end{kframe}
\end{knitrout}

\end{appendix}


\end{document}